\documentclass[a4paper]{jpconf}
\usepackage{graphicx} 
\usepackage{rotating}
\usepackage{amssymb}
\usepackage{amsmath}
\usepackage{mathptmx}
\usepackage{bm}

\begin{document}
\title{Closed-channel contribution in the BCS-BEC crossover regime of an ultracold Fermi gas with an orbital Feshbach resonance}
\author{S Mondal$^1$, D Inotani$^1$ and Y Ohashi$^1$}
\address{Department of Physics, Faculty of Science and Technology, Keio University, 3-14-1 Hiyoshi, Kohoku-ku, Yokohama 223-8522, Japan}
\ead{soumita\_phy@keio.jp}
\begin{abstract}
We theoretically investigate strong-coupling properties of an ultracold Fermi gas with an orbital Feshbach resonance (OFR). Including tunable pairing interaction associated with an OFR within the framework of the strong-coupling theory developed by Nozi\`eres and Schmitt-Rink (NSR), we examine the occupation of the closed channel. We show that, although the importance of the closed channel is characteristic of the system with an OFR, the occupation number of the closed channel is found to actually be very small at the superfluid phase transition temperature $T_{\rm c}$, in the whole BCS (Bardeen-Cooper-Schrieffer)-BEC (Bose-Einstein condensation) crossover region, when we use the scattering parameters for an ultracold $^{173}$Yb Fermi gas. The occupation of the closed channel increases with increasing the temperature above $T_{\rm c}$, which is more remarkable for a stronger pairing interaction. We also present a prescription to remove effects of an experimentally inaccessible deep bound state from the NSR formalism, which we meet when we theoretically deal with a $^{173}$Yb Fermi gas with an OFR.
\end{abstract}
\par
\section{Introduction}
\par
A magnetic Feshbach resonance (MFR) has proved to be a prerequisite for the study of an ultracold Fermi gas of alkali-metal atoms over a wide range of the strength of a pairing interaction, because it enables us to experimentally tune the interaction strength\cite{Chin}. One typical success by using this unique property is the experimental realization of the BCS (Bardeen-Cooper-Schrieffer)-BEC (Bose-Einstein condensation) crossover phenomenon\cite{NSR,Randeria,Ohashi} in $^{40}$K\cite{Jin} and $^6$Li\cite{Zwierlein} Fermi gases, where the character of a Fermi superfluid continuously changes from the weak-coupling BCS-type to the BEC of tightly-bound molecules, with increasing the interaction strength. 
\par
However, this MFR-technique is not applicable to the cases of alkali-earth atoms like Sr, as well as alkali-earth-like atoms like Yb, because of the vanishing total electronic spin due to the fully occupied outer-most shell. To overcome this problem, the possibility of using an optical Feshbach resonance (OFR) has been explored\cite{Enomoto}. However, this approach has serious problems, such as strong particle loss, as well as heating, so that it seems difficult to achieve the superfluid phase transition of a non-alkali metal Fermi gas by using this technique.
\par
Latest measurements showing the scattering properties of $^1$S$_0$ (electronic spin-singlet) and $^3$P$_0$ (spin-triplet) atomic pairs of $^{173}$Yb have predicted the existence of a shallow molecular bound state\cite{Bloch,Inguscio}. Very recently, a novel idea using the scattering between $^1$S$_0$ and $^3$P$_0$ in a $^{173}$Yb Fermi gas has been proposed\cite{Zhang}. This proposal uses different nuclear spin states of the atoms to tune the interaction strength, unlike a MFR that uses different electronic spin states. This so-called orbital Feshbach resonance (OFR) is expected to provide a tunable pairing interaction for an alkali-earth (like) atomic Fermi gas, where MFR cannot be deployed.  This OFR has recently experimentally been confirmed\cite{Inguscio2,Bloch2}. Measurements of anisotropic expansion of a gas cloud, as well as cross-thermalization rate, support the presence of OFR in an ultracold $^{173}$Yb Fermi gas at the field strengths $B=41$ G and 55(8) G. A relatively long lifetime of the system against collapse has also been reported\cite{Inguscio2,Bloch2}. Thus, the superfluid phase transition in an ultracold $^{173}$Yb Fermi gas with an OFR is promising.
\par
A crucial issue in theoretically considering an OFR is that the energy difference between the open and closed channels is comparable to the Fermi energy $\varepsilon_{\rm F}$ of the system or even smaller than $\varepsilon_{\rm F}$, so that we need to explicitly deal with the both channels\cite{Zhang}. This situation is quite different from the case of a broad (magnetic) Feshbach resonance in $^6$Li and $^{40}$K Fermi gases\cite{Chin,Jin,Zwierlein}, where the energy difference is so large that the close channel is only virtually occupied in the intermediate state of a Feshbach resonance. As a result, the latter can be treated by the simple single-channel BCS model, taking into account atoms only in the open channel. Effects of the closed channel only remains as a Feshbach-induced effective interaction between atoms in the open channel.
\par
In this paper, we theoretically investigate to what extent the closed channel contributes to the BCS-BEC crossover physics in an ultracold Fermi gas with an OFR. For this purpose, we evaluate the number of atoms in the closed channel within the framework of the strong-couping theory developed by Nozi\`eres and Schmitt-Rink (NSR)\cite{NSR}. We briefly note that the BCS-BEC crossover behavior of the superfluid phase transition temperature $T_{\rm c}$ in an ultracold Fermi gas with an OFR has been discussed\cite{Junjin} within the NSR theory\cite{NSR}, where higher superfluid transition temperature than the previous MFR case has been obtained in the weak-coupling regime. 
\par
\par
\section{Formulation}
\par
We consider a four-component Fermi gas with an OFR, described by the model Hamiltonian\cite{Zhang,Junjin},
\begin{equation}
H=K_{\rm o}+K_{\rm c}
+{1 \over 2}\sum_{\bm q}
\left[
U_+A_+^\dagger({\bm q})A_+(-{\bm q})
+
U_-A_-^\dagger({\bm q})A_-(-{\bm q})
\right].
\label{eq.1}
\end{equation}
Here, $K_{\rm o}=\sum_{\bm p}\xi_{\bm p}^{\rm o}[c_{{\rm g}\downarrow,{\bm p}}^\dagger c_{{\rm g}\downarrow,{\bm p}}+c_{{\rm e}\uparrow,{\bm p}}^\dagger c_{{\rm e}\uparrow,{\bm p}}]$ and $K_{\rm c}=\sum_{\bm p}\xi_{\bm p}^{\rm c}[c_{{\rm g}\uparrow,{\bm p}}^\dagger c_{{\rm g}\uparrow,{\bm p}}+c_{{\rm e}\downarrow,{\bm p}}^\dagger c_{{\rm e}\downarrow,{\bm p}}]$ are kinetic terms in the open channel ($|{\rm g}\downarrow\rangle$, $|{\rm e}\uparrow\rangle$) and closed channel ($|{\rm g}\uparrow\rangle$, $|{\rm e}\downarrow\rangle$), respectively. $\xi_{\bm p}^{\rm o}={\bm p}^2/(2m)-\mu$ is the energy in the open channel, and $\xi_{\bm p}^{\rm c}={\bm p}^2/(2m)+\delta/2-\mu$ is that in the closed channel, with $m$ being the atomic mass. In the above, ``g" and ``e" denote two different orbital states, and pseudospins $\sigma=\uparrow,\downarrow$ describe two nuclear spin states. $\delta/2$ is the energy difference between the open and closed channel. The last two terms in Eq. (\ref{eq.1}) describe interactions between Fermi atoms, where 
\begin{equation}
A_\pm({\bm q})=\sum_{\bm p}
\left[
c_{{\rm g}\uparrow,{\bm p}+{\bm q}/2} c_{{\rm e}\downarrow,-{\bm p}+{\bm q}/2}
\mp
c_{{\rm g}\downarrow,{\bm p}+{\bm q}/2} c_{{\rm e}\uparrow,-{\bm p}+{\bm q}/2}
\right].
\label{eq.2}
\end{equation}
The bare coupling strengths $U_\pm~(<0)$ are related to the observable $s$-wave scattering length $a_{s,\pm}$ as,
\begin{equation}
{4\pi a_{s,\pm} \over m}=
{U_\pm \over 1+U_\pm\sum_{\bm p}{m/{\bm p}^2}}.
\label{eq.3}
\end{equation}
\par
\begin{figure}[t]
\center
\includegraphics[width=0.73\textwidth]{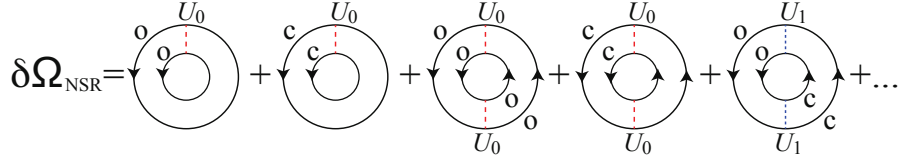}
\caption{Strong-coupling corrections ($\delta\Omega_{\rm NSR}$) to the thermodynamic potential in NSR. The solid line with label ``o" and ``c" represent the bare single-particle thermal Green's function $G_{{\rm s}={\rm o,c}}^{-1}({\bm p},i\omega_n)=i\omega_n-\xi_{\bm p}^{\rm s}$ in the open and closed channel, respectively, where $\omega_n$ is the fermion Matsubara frequency. The red and blue dashed line denote the intra- and inter-channel interactions, respectively. (Color figure online)}
\label{fig1}
\end{figure}

\par
Unlike a broad MFR (where the energy difference $\delta/2$ between the open and closed channels is much larger than the Fermi energy $\varepsilon_{\rm F}$), $\delta/2$ is comparable or even smaller than $\varepsilon_{\rm F}$ in the case of OFR\cite{Zhang}. Thus, we need to take into account scattering effects on both the open and the closed channels in the latter. In this paper, we include pairing fluctuations induced by the interactions $U_\pm$, within the framework of the strong-coupling (Gaussian fluctuation) theory developed by Nozi\`eres and Schmitt-Rink (NSR)\cite{NSR}. For this purpose, it is convenient to rewrite the interaction terms ($\equiv H_{\rm int}$) in Eq. (\ref{eq.1}) to the sum of intra-channel and inter-channel interactions, as
\begin{eqnarray}
H_{\rm int}
&=&
U^{\rm open}_{\rm intra}\sum_{{\bm p},{\bm p}',{\bm q}}
c_{{\rm e}\uparrow,{\bm p}+{\bm q}/2}^\dagger
c_{{\rm g}\downarrow,-{\bm p}+{\bm q}/2}^\dagger
c_{{\rm g}\downarrow,-{\bm p}'+{\bm q}/2}
c_{{\rm e}\uparrow,{\bm p}'+{\bm q}/2}
\nonumber
\\
&+&
U^{\rm closed}_{\rm intra}\sum_{{\bm p},{\bm p}',{\bm q}}
c_{{\rm g}\uparrow,{\bm p}+{\bm q}/2}^\dagger
c_{{\rm e}\downarrow,-{\bm p}+{\bm q}/2}^\dagger
c_{{\rm e}\downarrow,-{\bm p}'+{\bm q}/2}
c_{{\rm g}\uparrow,{\bm p}'+{\bm q}/2}
\nonumber
\\
&+&
U_{\rm inter}\sum_{{\bm p},{\bm p}',{\bm q}}
\left[
c_{{\rm e}\uparrow,{\bm p}+{\bm q}/2}^\dagger
c_{{\rm g}\downarrow,-{\bm p}+{\bm q}/2}^\dagger
c_{{\rm e}\downarrow,-{\bm p}'+{\bm q}/2}
c_{{\rm g}\uparrow,{\bm p}'+{\bm q}/2}
+{\rm h.c.}
\right].
\label{eq.4}
\end{eqnarray}
Here, $U_{\rm intra}^{\rm open}$, $U_{\rm intra}^{\rm closed}$, $U_{\rm inter}$ represent, respectively, the interaction in the open channel, that in the closed channel, and the inter-channel interaction. These are related to $U_\pm$ in Eq. (\ref{eq.1}) as $U_{\rm intra}^{\rm open}=U_{\rm intra}^{\rm closed}=[U_-+U_+]/2~(\equiv U_0)$ and $U_{\rm inter}=[U_--U_+]/2~(\equiv U_1)$. Summing up the NSR diagrams in Fig. \ref{fig1} describing fluctuation corrections ($\delta\Omega_{\rm NSR}$) to the thermodynamic potential $\Omega=\Omega_0+\delta\Omega_{\rm NSR}$, we obtain 
\begin{eqnarray}
\Omega = \Omega_0+ T
\sum_{{\bm q},\nu_n} \ln 
\left[
\det
\left[
\begin{array}{cc}
1-U_0\Pi_{\rm o}({\bm q},i\nu_n) & -U_1\Pi_{\rm o}({\bm q},i\nu_n) \\
-U_1\Pi_{\rm c}({\bm q},i\nu_n) & 1-U_0\Pi_{\rm c}({\bm q},i\nu_n)
\end{array}
\right]
\right],
\label{eq.5}
\end{eqnarray}
where $\Omega_0=-T\sum_{{\bm p},{\rm s}={\rm o,c}}\ln[1+e^{-\xi_{\bm p}^{\rm s}/T}]$ is the thermodynamic potential in the non-interacting case, and $\nu_n$ is the boson Matsubara frequency. In Eq. (\ref{eq.5}), 
\begin{equation}
\Pi_{{\rm s}={\rm o,c}}({\bm q},i\nu_n) = 
\sum_{\bm p}
{
1-f(\xi_{{\bm p}+{\bm q}/2}^{\rm s})-f(\xi_{-{\bm p}+{\bm q}/2}^{\rm s})
\over
i\nu_n-\xi_{{\bm p}+{\bm q}/2}^{\rm s}-\xi_{-{\bm p}+{\bm q}/2}^{\rm s}
}
\label{eq.6}
\end{equation}
is the lowest-order pair-correlation function, where $f(x)$ is the Fermi distribution function. 
\par
In the NSR scheme\cite{NSR}, the superfluid phase transition temperature $T_{\rm c}$ is determined from the Thouless criterion\cite{Thouless}, stating that the particle-particle scattering matrix has a pole in the low-energy and low-momentum limit. When we naively use the NSR thermodynamic potential in Eq. (\ref{eq.5}), this condition is achieved when ${\rm det}[\cdot\cdot\cdot]$ in Eq. (\ref{eq.5}) vanishes at ${\bm q}=\nu_n=0$, which gives,
\begin{equation}
1-U_0
[\Pi_{\rm o}(0,0)+\Pi_{\rm c}(0,0)]+(U_0^2-U_1^2)\Pi_{\rm o}(0,0)\Pi_{\rm c}(0,0)=0.
\label{eq.7}
\end{equation}
We then solve the $T_{\rm c}$-equation (\ref{eq.7}), together with the equation for the total number $N$ of Fermi atoms,
\begin{equation}
N=
-{\partial \Omega \over \partial \mu}=
\sum_{{\rm s}={\rm o.c}}N_{\rm s}
-
T\sum_{{\bm q},\nu_n}
{\partial \over \partial \mu}
\ln 
\left[
\det
\left[
\begin{array}{cc}
1-U_0\Pi_{\rm o}({\bm q},i\nu_n) & -U_1\Pi_{\rm o}({\bm q},i\nu_n) \\
-U_1\Pi_{\rm c}({\bm q},i\nu_n) & 1-U_0\Pi_{\rm c}({\bm q},i\nu_n)
\end{array}
\right]
\right]
\label{eq.8}
\end{equation}
(where $N_{\rm s}=2\sum_{\bm p}f(\xi_{\bm p}^{\rm s})$), to determine $T_{\rm c}$ and $\mu(T_{\rm c})$ self-consistently.
\par
However, when we apply this formalism to an ultracold $^{173}$Yb Fermi gas with an OFR ($a_{s,+}$=1900$a_0$ and $a_{s,-}$=200$a_0$, where $a_0=0.529~{\rm A}$ is the Bohr's radius\cite{Inguscio,Bloch2}), in addition to the experimentally accessible bound state with shallow binding energy $E_+^{\rm bind}$= -1/m$a_{s,+}^2$, one obtains another bound state which is experimentally inaccessible because of the very deep energy level $E_+^{\rm bind}=-1/ma_{s,-}^2\ll E_-$\cite{HuiHu}. Although the $T_{\rm c}$ equation (\ref{eq.7}) naively gives the Bose-Einstein condensation associated with the latter molecular state as the {\it highest} $T_{\rm c}$, we need to remove this ``experimentally inaccessible part" from the theory, in order to focus on the experimentally accessible shallow bound state.
\par
This attempt is achieved by diagonalizing the $2\times 2$ matrix part in Eq. (\ref{eq.5}) as 
\begin{eqnarray}
\Omega = \Omega_0+ T
\sum_{{\bm q},\nu_n} \ln 
\left[
\det
\left[
\begin{array}{cc}
\lambda_+({\bm q},i\nu_n) &0 \\
0& \lambda_-({\bm q},i\nu_n)
\end{array}
\right]
\right],
\label{eq.9}
\end{eqnarray}
where
\begin{equation}
\lambda_\pm({\bm q},i\nu_n) =
1-{U_0 \over 2}[\Pi_{\rm o}+\Pi_{\rm c}]
\pm
{1 \over 2}
\sqrt{U_0^2[\Pi_{\rm o}^2+\Pi_{\rm c}^2]-2[U_0^2-2U_1^2]\Pi_{\rm o}\Pi_{\rm c}}.
\label{eq.10}
\end{equation}
Between $\lambda_\pm$, one finds that $\lambda_+$ corresponds to the accessible shallow state. Only retaining this contribution, we obtain the number equation, as well as the $T_{\rm c}$-equation as, respectively,
\begin{equation}
N=\sum_{{\rm s}={\rm o,c}}N_{\rm s}
-T\sum_{{\bm q},\nu_n}
{\partial \over \partial\mu}\ln[\lambda_+({\bm q},i\nu_n)],
\label{eq.11}
\end{equation}
\begin{equation}
1={U_0 \over 2}[\Pi_o(0,0)+\Pi_{\rm c}(0,0)]
+{1 \over 2}
\sqrt{U_0^2[\Pi_{\rm o}^2(0,0)+\Pi_{\rm c}^2(0,0)]-2[U_0^2-2U_1^2]\Pi_{\rm o}(0,0)\Pi_{\rm c}(0,0)}.
\label{eq.12}
\end{equation}
We briefly note that we only deal with the number equation (\ref{eq.11}), when we evaluate $\mu(T>T_{\rm c})$.
\par
In contrast to the case of a broad magnetic Feshbach resonance, where the interaction effects are simply tuned by the scaled inverse $s$-wave scattering length $(k_{\rm F}a_s)^{-1}$ (where $k_{\rm F}$ is the Fermi momentum), the present OFR case is known to also depend on other parameters, such as the number density $n$\cite{Junjin}. In this paper, we set the number density as $n=k_{\rm F}^3/(3\pi^2)=5\times 10^{13} {\rm cm}^{-3}$\cite{Zhang,Junjin}. We also take the energy difference $\delta/2$ between the open and closed channel as $\delta/2=(2\pi\hbar\times 56\Delta_m)B$ Hz\cite{Zhang,Junjin} (where $\Delta_m$=5 is the difference between the real nuclear-spin quantum numbers described by pseudospin $\sigma=\uparrow,\downarrow$, and $B$~[G] is an external magnetic field). As in the MFR case, we measure the interaction strength in terms of the scaled inverse $s$-wave scattering length $(k_{\rm F}a_s)^{-1}$ in the open channel, where $a_s$ is given by\cite{Zhang}
\begin{equation}
a_s=
{
-a_{s0}+\sqrt{m\delta/\hbar^2}[a_{s0}^2-a_{s1}^2]
\over
a_{s0}\sqrt{m\delta/\hbar^2}-1}~~~(\delta>0),
\label{eq.13}
\end{equation}
where $a_{s0}=[a_{s,+}+a_{s,-}]/2$ and $a_{s1}=[a_{s,-}-a_{s,+}]/2$.
\par
\begin{figure}[t]
\center
\includegraphics[width=0.79\textwidth]{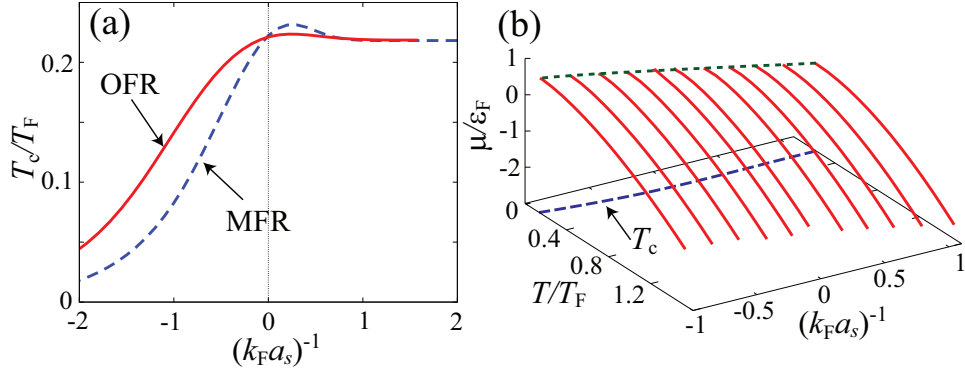}
\caption{(a) Calculated $T_{\rm c}$ in an ultracold Fermi gas with an OFR, as a function of the interaction strength  $(k_{\rm F}a_s)^{-1}$ . The dashed line shows the result in the MFR case. (b) Fermi chemical potential $\mu(T\ge T_{\rm c})$ in the BCS-BEC crossover region. The dashed line shows $\mu(T_{\rm c})$. (Color figure online)}
\label{fig2}
\end{figure}
\par
\section {Results}
\par
Figure \ref{fig2}(a) shows the superfluid phase transition temperature $T_{\rm c}$, obtained from the coupled equations (\ref{eq.11}) with (\ref{eq.12}). ($\mu(T_{\rm c})$ is shown as the dotted line in Fig. \ref{fig2}(b).) Our result agrees well with the previous work\cite{Junjin}, confirming that our prescription for removing inaccessible deep bound state really works. As pointed out in Ref.\cite{Junjin}, OFR gives higher $T_{\rm c}$ in the weak-coupling side, compared to the case of a broad magnetic Feshbach resonance (MFR).
\par
\begin{figure}[t]
\center
\includegraphics[width=0.72\textwidth]{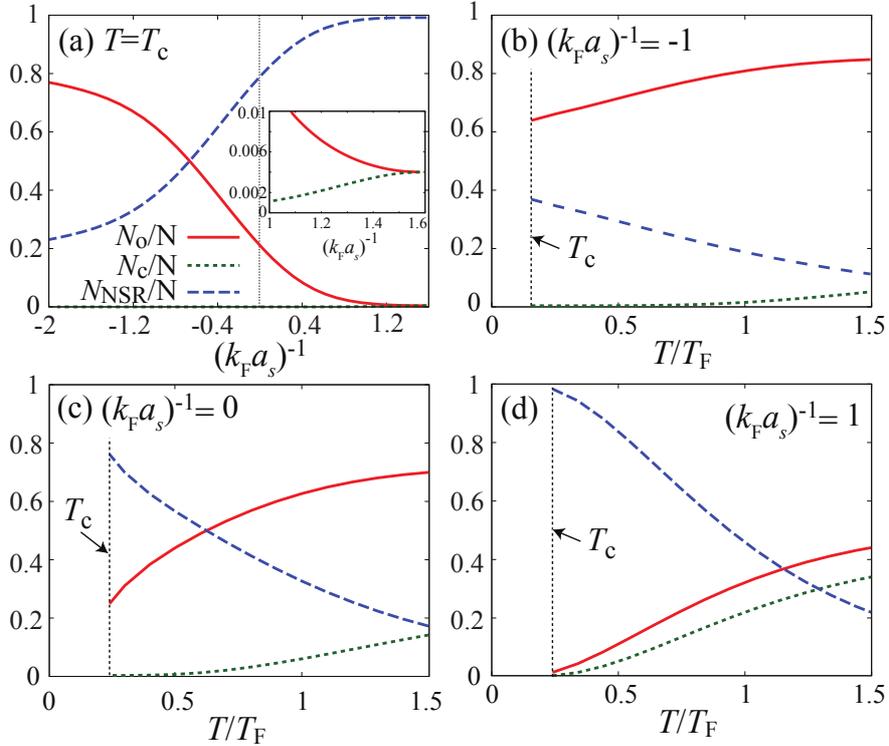}
\caption{(a) The number of atoms in the open channel $N_{\rm o}=\sum_{\bm p}f(\xi_{\bm p}^{\rm o})$, as well as that in the closed channel $N_{\rm c}=\sum_{\bm p}f(\xi_{\bm p}^{\rm c})$ in the BCS-BEC crossover region at $T_{\rm c}$. $N_{\rm NSR}=N-N_{\rm o}-N_{\rm c}$ describes NSR fluctuation contribution. For clarity, we magnify the result in the strong-coupling regime in the inset. (b)-(d) Temperature dependence of $N_{\rm o}$, $N_{\rm c}$, and $N_{\rm NSR}$: (b) weak-coupling regime. (c) unitarity limit. (d) strong-coupling regime. (Color figure online)}
\label{fig3}
\end{figure}

To see to what extent atoms in the closed channel contribute to the superfluid instability, we plot the number $N_{\rm c}$ of atoms at $T_{\rm c}$ in Fig. \ref{fig3}(a). Then, we find that, although the necessity of explicit treatment of the closed channel is characteristic of OFR, $N_{\rm c}(T=T_{\rm c})$ is actually very small in the entire BCS-BEC crossover region. $N_{\rm c}$ reaches the number $N_{\rm o}$ of atoms in the open channel, when the energy difference $\delta/2$ between the open and closed channels vanishes at $(k_{\rm F}a_s)^{-1}=1.58$ (see the inset in Fig. \ref{fig3}(a)). However, this strong-coupling regime is already dominated by tightly bound molecules described by the ``NSR" term $N_{\rm NSR}=N-N_{\rm o}-N_{\rm c}$, as shown in Fig. \ref{fig3}(a). Thus, in the case of a $^{173}$Yb Fermi gas, the system properties at $T_{\rm c}$ are dominated by the open channel ($N_{\rm o}$) and fluctuation contribution ($N_{\rm NSR}$), as in the case of a broad magnetic Feshbach resonance used in $^6$Li and $^{40}$K Fermi gases.
\par
\par
The occupation of the closed channel gradually increases with increasing the temperature above $T_{\rm c}$, as shown in Figs. \ref{fig3}(b)-(d). Here, these results are obtained by using the Fermi chemical potential $\mu(T\ge T_{\rm c})$ shown in Fig. \ref{fig2}(b), calculated from the number equation (\ref{eq.11}). Since the energy difference $\delta/2$ between the open and closed channels is smaller for a stronger pairing interaction (see Eq. (\ref{eq.13})), the growth of $N_{\rm c}$ is more remarkable in the strong-coupling side (Fig. \ref{fig3}(d)) than in the weak-coupling case (Fig. \ref{fig3}(b)). We briefly note that, in all the cases shown in Figs. \ref{fig3}(b)-(d), both $N_{\rm o}$ and $N_{\rm c}$ monotonically increase and the fluctuations contribution $N_{\rm NSR}$ only decreases, as one increases the temperature. This means that the atoms in the closed channel ($N_{\rm c}$) are supplied by the decrease of atoms in the scattering states, as well as by the dissociation of tightly bound molecules (that are described by $N_{\rm NSR}$), rather than thermal excitations from the open channel to the closed channel. We briefly note that the closed-channel band $\xi_{\bm p}^{\rm c}={\bm p}^2/(2m)+\delta/2-\mu$ is higher than the open-channel band $\xi_{\bm p}^{\rm o}={\bm p}^2/(2m)-\mu$ by $\delta/2>0$. Thus, when the temperature is much higher than this energy difference ($T\gg \delta/2$), the number $N_{\rm c}$ of atoms in the closed would approach that in the open channel as $N_{\rm o}\simeq N_{\rm c}\simeq N/2$.
\par
\par
\section {Conclusion}
\par
To summarize, we have discussed an ultracold Fermi gas with an orbital Feshbach resonance (OFR). Including both the open and closed channels, as well as pairing fluctuations, within the framework of the strong-coupling theory developed by Nozi\`eres and Schmitt-Rink (NSR), we have calculated the numbers of atoms in the open channel ($N_{\rm o}$) and closed channel ($N_{\rm c}$), as well as the NSR fluctuation contribution ($N_{\rm NSR}=N-N_{\rm o}-N_{\rm c}$), in the BCS-BEC crossover region above $T_{\rm c}$. In this procedure, we have also presented an idea to remove effects of the experimentally inaccessible bound state which appears when a $^{173}$Yb Fermi gas is considered.
\par
We showed that, when we take parameters for a $^{173}$Yb Fermi gas, the occupation of the closed channel is actually very small over the entire BCS-BEC crossover region at $T_{\rm c}$. On the other hand, the occupation number of the closed channel gradually becomes remarkable, as one raises the temperature above $T_{\rm c}$, especially in the strong-coupling regime. This result implies that atoms in the closed channel (which is characteristic of the OFR case) affect the temperature dependence of physical quantities. Since the superfluid phase transition in a $^{173}$Yb Fermi gas with an OFR is promising, our results would contribute to the research toward the realization of a Fermi superfluid of non-alkali metal atomic gas.
\par
\section* {Acknowledgements}
We thank P. van Wyk, D. Kharga and R. Hanai for useful discussions. This work was supported by the KiPAS project in Keio university. DI was supported by Grant-in-Aid for Young Scientists (B) (No. JP16K17773) from JSPS. YO was supported by Grand-in-Aid for Scientific Research from MEXT and JSPS in Japan (No. JP15K00178, No. JP15H00840, No. JP16K05503).
\par
\par

\end{document}